\documentclass[10pt]{article} 

\newcommand{\m}{\mathbf}
\newcommand{\be}{\begin{equation}}
\newcommand{\ee}{\end{equation}}
\newcommand{\ba}{\begin{eqnarray}}
\newcommand{\ea}{\end{eqnarray}}
\newcommand{\nn}{\nonumber}

\newcommand{\ra}{\rightarrow}

\renewcommand{\lim}{\textrm{lim\,}}

\newcommand{\dd}{\textrm{d}}

\newcommand{\R}{\mathrm{R}}

\newcommand{\tr}{\textrm{tr\,}}
\newcommand{\Tr}{\textrm{tr$_\O$\,}}

\newcommand{\s}{\,\,\,}

\newcommand{\fr}{\frac}

 \renewcommand{\d}{\delta}
\newcommand{\e}{\epsilon} 
 \renewcommand{\m}{\mu} \newcommand{\n}{\nu}
\renewcommand{\r}{\rho} \renewcommand{\s}{\sigma}

\renewcommand{\O}{\Omega} 
 
\renewcommand{\R}{\mathsf{R}} 
\renewcommand{\O}{\mathsf{O}}

\newcommand{\bsm}{\begin{smallmatrix}}
\newcommand{\esm}{\end{smallmatrix}} 

\renewcommand{\star}{\diamond}

\begin{document}

\title{ Construction of multi-instantons in \\  eight dimensions}

\author{Roman V. Buniy and Thomas W. Kephart\\ \emph{Department of
Physics and Astronomy}\\ \emph{Vanderbilt University, Nashville, TN
37235}}

\date{3 October 2002}

\maketitle
             
\begin{abstract}

We consider an eight-dimensional local octonionic theory with the
seven-sphere playing the role of the gauge group. Duality conditions
for two- and four-forms in eight dimensions are related. Dual
fields---octonionic instantons---solve an 8D generalization of the
Yang-Mills equation. Modifying the ADHM construction of 4D instantons,
we find general $k$-instanton 8D solutions which depends on $16k-7$
effective parameters.

\end{abstract}

\renewcommand{\thefootnote}{\arabic{footnote}}
\setcounter{footnote}{0}

\section{Introduction.}

The discovery of instantons~\cite{instanton} was an important advance
in our understanding of non-perturbative quantum field theory. These
objects are (anti-)self-dual ($*F=\pm F$) Euclidean solutions to
Yang-Mills field equations in 4D. They have lead to a deeper
understanding of the QCD vacuum ($\theta$ vacuum~\cite{vacuum}), and
have been conjectured to play a part in the confinement of color
charges~\cite{confinement}. Instantons also have a broad significance
in mathematics, specifically in the theory of fake
$\R^4$-manifolds~\cite{donaldson}. The most general multi-instanton
solutions have been constructed~\cite{Atiyah:1}, and these again
played a part in broadening our understanding of gauge theories.

A single instanton solution is spherically symmetric and, in
mathematical language, corresponds to the third Hopf map, which is the
principal fibre bundle $S^7\stackrel{S^3}{\ra}S^4$, where $S^4$ is the
one-point compactification of $\R^4$, $S^3\sim SU(2)$ is the fibre
(gauge group) and $S^7$ is the total space.

As string theory and M-theory live in higher dimensions, it is of
interest to consider higher dimensional analogs of 4D instantons; in
particular, there exists a natural generalization of instantons to 8D,
where the last Hopf map $S^{15}\stackrel{S^7}{\ra}{S^8}$ resides. The
original 4D instanton had gauge group $SU(2)$ embedded in $Spin(4)\sim
SU(2)\times SU(2)$, so that the bundle became
$Spin(5)\stackrel{Spin(4)}{\longrightarrow}S^4$. The analogous single
instanton solution in 8D was found in \cite{Grossman}, and has a
generalized self-duality $*F^2=\pm F^2$ with the bundle
$Spin(9)\stackrel{Spin(8)}{\longrightarrow}S^8$. The higher
dimensional instanton have conformal features similar to those of 4D
instantons. The 8D case, and especially its multi-instanton
generalization, appears more complicated than its 4D counterpart for
the following reasons:
\begin{enumerate} \item{The fibre that is twisted with the 4D base
space is a three-sphere, but this is a group, while the twisted part
of the $Spin(8)\sim S_L^7\times S_R^7\times G_2$ fibre is a
seven-sphere. $S^7$ is the only paralellizable manifold that is not a
Lie group, but it does have a close resemblance to a gauge group.}
\item{As $S^7$ can be represented by unit octonions, and $G_2$ is the
automorphism group of the octonions, there is a hidden
nonassociativity that comes into play.}
\item{There is only one choice (via the Hodge star) for the form of
duality in 4D, but in 8D other possibilities arise, e.g., a tensor
form of duality $\lambda F_{\m\n}=\fr{1}{2}T_{\m\n\r\s}F_{\r\s}$ has
been studied~\cite{duality,O-instanton}.}
\end{enumerate}

Attempts \cite{Tchrakian} to obtain multi-instantons in a $Spin(8)$
gauge theory meet with a number of difficulties. To circumvent these
obstructions, we turn to a theory with only $S^7$ fibre, but to do
this, we first need to review the properties of the octonions. Here we
will construct multi-instanton solutions in 8D through a
generalization of the ADHM procedure, and to do this we must deal with
all of the above complications. We will introduce products and
operators in a way that nonassociativity is tamed. Next, a new
generalized duality is used to provide results that allow us to relate
the topologically significant quadratic duality on $F^2$ to a specific
form of tensor duality. We then consider the symmetries of our
multi-instanton solutions and show that in 8D the $k$-instanton $S^7$
bundles contain $16k-7$ parameters in analogy with the $8k-3$
parameters of the most general 4D $k$-instanton $S^3$ bundles.

\section{Octonions}

We recall (for a review, see e.g. \cite{octonions}) that the
nonassociative octonionic algebra has the multiplication rule $e_i
e_j=-\d_{ij}+f_{ijk}e_k$, where the $f_{ijk}$'s are completely
anti-symmetric structure constants. The seven-sphere is described by a
unit octonion $g$ satisfying $g^*g=1$. The octonions' nonassociativity
complicates construction of the analog of a gauge theory. For example,
for imaginary octonionic $A$ and $F=\dd A+A^2$, the corresponding
$S^7$-gauge transformed quantities are \ba A_g=g^*Ag+g^*\dd g\nn\ea
and \ba&&F_g=g^*Fg\nn\\&&\ \ \ \ \ +\dd g^*\dd g-(\dd g^*g)(g^*\dd
g)+\dd g^*(Ag)-(\dd g^*g)(g^*Ag)\nn\\ &&\ \ \ \ \ -g^*(A\dd
g)+(g^*Ag)(g^*\dd g)-g^*A^2g+(g^*Ag)(g^*Ag).\label{Fg}\ea
Nonassociativity prevents the terms in the last two lines of
(\ref{Fg}) from canceling. Using $g(g^*h)=h$, which holds for any
octonions $g$ and $h$, we note that the terms do cancel in $LF_g$,
where $L$ is the operator of left octonionic multiplication, \ba
&&L(a_1\ldots a_n)=a_1(a_2(a_3(\ldots a_n))\ldots).\label{L}\ea Any
arrangement of parentheses in the argument of $L$ give the same
results on the right-hand side of (\ref{L}). Use of the operator $L$
allows us to perform various operations on the octonions as if they
were associative. For simplicity in notations, we omit $L$ in the
following. Instead of left octonionic multiplication we could use
right multiplication with the same result. From (\ref{Fg}) we now find
the familiar result $F_g=g^*Fg$.

For associative $A$ and $F$, the forms $\tr F^n$ are closed. To extend
this to octonions, which do not admit a matrix representation, we need
an octonionic operator with some of the properties of the matrix
trace. Consider the operator $\Tr$ defined by \ba\Tr L(a_1\ldots
a_n)=\fr{1}{n}\sum_{k=1}^n
(-1)^{(r_k+\ldots+r_n)(r_1+\ldots+r_{k-1})}L(a_k\ldots a_na_1\ldots
a_{k-1}),\label{trace}\ea where differential forms $a_k$ are of
degrees $r_k$. The operators $\Tr$ and $\dd$ commute and so the forms
$\Tr F^n$ are closed; thus we arrive at the familiar Lie algebra
result~\cite{Eguchi}: \ba\Tr F^n=\dd Q_{2n-1},\label{dQ}\ea where \ba
Q_{2n-1}=n\Tr\int_0^1\dd t
A\left[tF+(t^2-t)A^2\right]^{n-1}.\label{Q}\ea

\section{Linear duality}

Since any pair of imaginary octonions generate a quaternionic
subalgebra, we expect to find an octonionic duality condition which is
reducible to its quaternionic counterpart. For example, let us define
dual octonionic $2$-forms according to \ba&&\star(\dd x_\m\dd
x_\n)=\fr{1}{2}f_{\m\n\r\s}\dd x_\r\dd x_\s,\label{star}\ea and
determine the tensor $f_{\m\n\r\s}$ from the following two
requirements: (i) any $2$-form can be written as a sum of its
self-dual and anti-self-dual parts, or equivalently, $\star^2=1$; (ii)
$\dd x\dd x^*$ is self-dual and $\dd x^*\dd x$ is
anti-self-dual. Consequently, for octonionic forms we obtain
\ba&&f_{0ijk}=f_{ijk},\nn\\&&f_{ij0k}=\fr{1}{3}f_{ijk},\nn\\
&&f_{ijkl}=\pm\fr{1}{3}f_{ijm}f_{klm}\mp(\d_{ik}\d_{jl}
-\d_{il}\d_{jk}).\label{f}\ea

From Eqs.~(\ref{star}) and (\ref{f}), the components of the
$\star$-dual field strength $F=\fr{1}{2}F_{\m\n}\dd x_\m\dd x_\n$ are
subject to the following 21 relations:~\footnote{While our octonionic
duality condition (\ref{dual}) is similar in form to one of the two
duality conditions for $SO(8)$ considered in Ref.~\cite{duality}, the
latter were not constructed to satisfy either of the two
above-mentioned requirements. Consequently, our octonionic instantons
are different from the $SO(8)$ solutions in Ref.~\cite{O-instanton}.}
\ba F_{ij}=\pm f_{ijk}F_{0k}.\label{dual}\ea

Applied to the quaternions, the above requirements lead to the
familiar relations $f_{0abc}=f_{ab0c}=\e_{abc}$ and $f_{abcd}=0$. In
both the quaternionic and octonionic cases, the components
$f_{\m\n\r\s}$ are the matrix elements of the corresponding groups and
cosets in the products $Spin(4)=S^3_L\times S^3_R$ and
$Spin(8)=S^7_L\times G_2\times S^7_R$. Also, the components turn out
to coincide with the elements of the torsion and curvature tensors of
$Spin(4)/Spin(3)$ and $Spin(7)/G_2$ respectively (for the latter see
\cite{Lukierski}). Note the two choices of sign for the curvature
tensor $f_{\m\n\r\s}$ in (\ref{f}) and the two choices of orientation,
$S^7_{L,R}=Spin(7)_{L,R}/G_2$. Neither corresponds to the two choices
of sign in Eq.~(\ref{dual}).

Dual fields satisfy $\star F=\pm F$ and, in view of the octonionic
Bianchi identity $DF=0$, they also solve an 8D generalization of the
Yang-Mills equation $D\star F=0$. Below we find multi-particle
solutions to the duality equations.

\section{Quadratic duality}

In addition to the linear form of duality considered above, a
quadratic form of duality is also possible in 8D. In the latter case,
dual octonionic 4-forms are related via the Hodge star, ``$*$''.

A conformally invariant action $I=\Tr\int F^2*F^2$ yields the equation
of motion $\{F,D*F^2\}=0$. The $*$-dual fields, which are defined by
\ba *F^2=\pm F^2,\label{FF}\ea solve the equation of motion by means
of the Bianchi identity $DF=0$.

In terms of (anti-)self-dual $F^2_\pm=\fr{1}{2}(F^2\pm *F^2)$, the
action becomes \ba I=\Tr\int\left(F^2_+*F^2_++F^2_-*F^2_-\right).
\label{I}\ea On the other hand, the topological charge (the forth 
Chern number) is \ba N=\frac{1}{384\pi^4}\,\Tr\int
F^4=\frac{1}{384\pi^4}\,\Tr\int(F^2_+*F^2_+-F^2_-*F^2_-),\label{k}\ea
where we have used $F^2_\pm F^2_\mp=0$. It follows from (\ref{I}) and
(\ref{k}) that the action is bounded from below, \ba I\ge
384\pi^4|n|,\nn\ea with minima achieved when $F^2_\pm=0$, i.e. for the
$*$-dual fields (\ref{FF}). There are one-particle solutions to the
quadratic duality equations (\ref{FF}), and these solutions have a
geometric interpretation in terms of the forth Hopf
map~\cite{Grossman}.

It is remarkable but straightforward to verify that $\star F=\pm F$
implies $*F^2=\mp F^2$. To check this, we need the identity
\ba{\d^{\{i}}_{[k}{f^{j\}}}_{lm]}
=-\fr{1}{24}\e_{klmnpqr}{f^i}_{np}{f^j}_{qr},\nn\ea where indices
included in braces (brackets) are to be symmetrized
(antisymmetrized). We can also view $*$ as a ``square'' of
$\star$. The relation between the linear and quadratic dualities
allows us to proceed with the construction of octonionic
multi-instantons.

\section{Solution}

The ADHM construction~\cite{Atiyah:1} gives the most general
multi-instanton solutions to the duality equations in four
dimensions. We construct octonionic dual fields by a suitable 8D
generalization of the ADHM formula. Namely, consider a gauge
potential~\cite{Atiyah:2} \ba A=\frac{U^\dag\dd U-\dd U^\dag
U}{2(1+U^\dag U)},\ \ \ \ \ U^\dag=V(B-xI)^{-1},\label{A}\ea where the
$k$-dimensional vector $V$ and the $k\times k$ matrix $B$ have
constant octonionic entries. The operator $L$ is suppressed as usual,
and the symbol ``$\dag$'' means matrix transposition combined with
octonionic conjugation. The corresponding field strength is \ba
F=(1+U^\dag U)^{-2}U^\dag\dd xW\dd x^*U,\label{F}\ea where
$W^{-1}=V^\dag V+(B^\dag-x^*I)(B-xI)$.

For real $W$, i.e. when \ba&&V^\dag V+B^\dag B\ \ \textrm{is
real}\label{VB}\\&&\textrm{and}\ \ B\ \ \textrm{is symmetric},\nn\ea
$F$ involves the expression $L(\ldots\dd x\dd x^*\ldots)$. The
$\star$-dual of this 2-form is $L(\ldots\star(\dd x\dd x^*)\ldots)$
and, owing to the self-duality of $\dd x\dd x^*$, $F$ is
$\star$-self-dual itself, but $F^2$ is
$*$-anti-self-dual. Interchanging $x$ and $x^*$, interchanges
self-dual and anti-self-dual objects for both dualities.

\section{Instanton number}

For the solution obtained above, the gauge potential vanishes at
infinity faster than a pure gauge, and has singularities at the
instanton locations. A physically acceptable solution results from a
suitable gauge transformation.

The singularities are located at eigenvalues $\{b_i\}$ of the $k\times
k$ matrix $B$. Expanding around each singularity, we have
approximately \ba A\approx\frac{y_i^*\dd y_i-\dd y_i^*y_i}
{2|y_i|^2(1+|y_i|^2)}\ \ \ \textrm{for}\ \ \ y_i\to 0,\label{Ai}\ea
where $y_i=(x-b_i)V_i^*$. A gauge transformation with the gauge
function $g_i=y_i^*/|y_i|$ removes the singularity at $y_i=0$ in the
potential (\ref{Ai}), and leads to \ba A_{g_i}\approx\frac{y_i\dd
y_i^*-\dd y_i y_i^*} {2(1+|y_i|^2)}\ \ \ \textrm{for}\ \ \ y_i\to
0.\ea

Similar to the quaternionic case~\cite{Giambiagi}, all singularities
inside a finite $S^7$ can be removed. Inside this $S^7$, after using
(\ref{dQ}), the instanton number becomes \ba N=\int_{\R^8}\Tr
F_g^4=\int_{S^7}(Q_7)_g,\ea where asymptotically
$(Q_7)_g\sim-\frac{1}{35}\Tr(g^*\dd g)^7$. Since the field strength
corresponding to the gauge potential $g^*\dd g$ is zero, we use
Stokes's theorem again to replace the integral over the large $S^7$ by
the sum of the integrals over $k$ small spheres $S_i^7$ enclosing
singularities $b_i$. Around each singularity, $F_g$ looks like the
field of a single anti-instanton at the origin, \ba F_g=\frac{\dd x\dd
x^*}{(1+|x|^2)^2}.\label{F1}\ea Therefore, the topological charge $N$
and minus the instanton number $-k$ are one and the same.

\section{Parameters}

We now count the number of parameters needed to describe a
$k$-instanton. The octonions $V$ and $B$ have, respectively, $8k$ and
$8\fr{1}{2}k(k+1)$ real parameters. There are $7\fr{1}{2}k(k-1)$ real
equations in (\ref{VB}) constraining $V$ and $B$. When $V$ is replaced
by $g^*V$, where $g\in S^7$ is constant, the potential (\ref{A}) is
gauge transformed, $A\ra g^*Ag$, eliminating $7$ more
parameters. Also, a transformation $V\ra VT$, $B\ra T^{-1}BT$ with
real and constant $T\in O(k)$, which has $\fr{1}{2}k(k-1)$ parameters,
does not change $A$. Therefore, the number of effective degrees of
freedom desrcibing a $k$-instanton is \ba
8k+8\fr{1}{2}k(k+1)-7\fr{1}{2}k(k-1)-7-\fr{1}{2}k(k-1 )=16k-7.\ea We
do not have a proof that the above construction gives all dual fields,
although we suspect it does. At least it does so for the case of a
one-instanton~\cite{Grossman}, which is described by $9$
parameters---instanton's scale and location. Perhaps completeness of
the construction can be proved by using octonionic projective spaces
\cite{OP} and generalized twistors in analogy with the 4D case
(\cite{Atiyah:1,Atiyah:2}). Other multi-instanton solutions are
subsets of our solutions. For example, one can generalize Witten's and
't Hooft's~\cite{instantons} 4D multi-instanton solutions to 8D.

The single 8D instanton has entered string theory and produced a
solitonic member of the brane scan (for a review, see \cite{Duff}). We
hope our general construction will facilitate further applications to
string and M-theory, and perhaps in pure mathematics.


\begin{thebibliography}{99}


\bibitem{instanton} A.~A.~Belavin, A.~M.~Polyakov, A.~S.~Schwartz and
Y.~S.~Tyupkin, Phys.\ Lett.\ B {\bf 59} (1975) 85.


\bibitem{vacuum} G.~'t Hooft, Phys.\ Rev.\ Lett.\ {\bf 37} (1976)
8;\newline C.~G.~Callan, R.~F.~Dashen and D.~J.~Gross, Phys.\ Lett.\ B
{\bf 63} (1976) 334;\newline R.~Jackiw and C.~Rebbi, Phys.\ Rev.\
Lett.\ {\bf 37} (1976) 172.


\bibitem{confinement} C.~G.~Callan, R.~F.~Dashen and D.~J.~Gross,
Phys.\ Rev.\ D {\bf 17} (1978) 2717.


\bibitem{donaldson} S.~K.~Donaldson and P.~B.~Kronheimer, The Geometry
of Four-Manifolds (Clarendon Press, Oxford, 1990).

\bibitem{Atiyah:1} M.~F.~Atiyah, N.~J.~Hitchin, V.~G.~Drinfeld and
Y.~I.~Manin, Phys.\ Lett.\ A {\bf 65} (1978) 185.

\bibitem{Grossman} B.~Grossman, T.~W.~Kephart and J.~D.~Stasheff,
Commun.\ Math.\ Phys.\ {\bf 96} (1984) 431 [Erratum-ibid.\ {\bf 100}
(1985) 311].

\bibitem{duality} E.~Corrigan, C.~Devchand, D.~B.~Fairlie and
J.~Nuyts, Nucl.\ Phys.\ B {\bf 214} (1983) 452;\newline D.~B.~Fairlie
and J.~Nuyts, J.\ Math.\ Phys.\ {\bf 25} (1984) 2025.

\bibitem{O-instanton} D.~B.~Fairlie and J.~Nuyts, J.\ Phys.\ A {\bf
17} (1984) 2867;\newline S.~Fubini and H.~Nicolai, Phys.\ Lett.\ B
{\bf 155} (1985) 369.

\bibitem{Tchrakian} J.~Spruck, D.~H.~Tchrakian and Y.~Yang, Commun.\
Math.\ Phys.\ {\bf 188} (1997) 737;\newline D.~H.~Tchrakian and
A.~Chakrabarti, J.\ Math.\ Phys.\ {\bf 32} (1991) 2532.

\bibitem{octonions} R.~D.~Schafer, An Introduction to Nonassociative
Algebras, (Academic Press, New York, 1966);\newline F.~G\"{u}rsey and
C.~H.~Tze, On the Role of Division, Jordan and Related Algebras in
Particle Physics, (World Scientific, Singapore, 1996).

\bibitem{Eguchi} T.~Eguchi, P.~B.~Gilkey and A.~J.~Hanson, Phys.\
Rept.\ {\bf 66} (1980) 213.

\bibitem{Lukierski} J.~Lukierski and P.~Minnaert, Phys.\ Lett.\ B {\bf
129} (1983) 392.

\bibitem{Atiyah:2} Michael Atiyah, Collected Works, (Clarendon Press,
Oxford, 1988), vol. 5, pp. 75--173.

\bibitem{Giambiagi} J.~J.~Giambiagi and K.~D.~Rothe, Nucl.\ Phys.\ B
{\bf 129} (1977) 111.

\bibitem{OP} M.~Cederwall and C.~R.~Preitschopf,
arXiv:hep-th/9403028;\newline J.~C.~Baez, arXiv:math.ra/0105155.

\bibitem{instantons} E.~Witten, Phys.\ Rev.\ Lett.\ {\bf 38} (1977)
121;\newline G.~'t~Hooft, unpublished;\newline R.~Jackiw, C.~Nohl and
C.~Rebbi, Phys.\ Rev.\ D {\bf 15} (1977) 1642.

\bibitem{Duff} M.~J.~Duff, R.~R.~Khuri and J.~X.~Lu, Phys.\ Rept.\
{\bf 259} (1995) 213 [arXiv:hep-th/9412184].








\end{thebibliography}
\end{document}